\documentclass[aps,prd,preprint,floatfix,nobibnotes,nofootinbib]{revtex4-2}

\usepackage{amsmath}
\usepackage{amssymb}
\usepackage{graphicx}
\usepackage{float}
\usepackage{dsfont}
\usepackage{ulem}
\usepackage[colorlinks=true]{hyperref}

\newcommand{\be}{\begin{equation}}
\newcommand{\ee}{\end{equation}}
\newcommand{\ba}{\begin{eqnarray}} 
\newcommand{\ea}{\end{eqnarray}} 
\newcommand{\nn}{\nonumber}
\newcommand{\bea}{\begin{eqnarray}}
\newcommand{\eea}{\end{eqnarray}}

\usepackage[usenames, dvipsnames]{color}

\numberwithin{equation}{section}

\begin{document} 

\title{Hydrodynamic-like behaviour of glasma}

\author{Margaret E. Carrington}
\affiliation{Department of Physics, Brandon University,
Brandon, Manitoba R7A 6A9, Canada}
\affiliation{Winnipeg Institute for Theoretical Physics, Winnipeg, Manitoba, Canada}

\author{Stanis\l aw Mr\' owczy\' nski} 
\affiliation{National Centre for Nuclear Research, ul. Pasteura 7,  PL-02-093 Warsaw, Poland}
\affiliation{Institute of Physics, Jan Kochanowski University, ul. Uniwersytecka 7, PL-25-406 Kielce, Poland}

\author{Jean-Yves Ollitrault}
\affiliation{Institut de Physique Th\'eorique, Universit\'e Paris Saclay, CNRS, CEA, F-91191 Gif-sur-Yvette, France} 

\date{November 8, 2024}

\begin{abstract}

At the earliest stage of ultrarelativistic heavy-ion collisions the produced matter is a highly populated system of gluons called glasma which can be approximately described in terms of classical chromodynamic fields. Although the system's dynamics is governed by Yang-Mills equations,  glasma evolution is shown to strongly resemble hydrodynamic behaviour. 

\end{abstract}

\maketitle
\newpage

%%%%%%%%%%%%%%%%%%%%%%%%%%%%%%%%%%%%%%%%%%%%%%%%%%%%%%%%
\section{Introduction}
\label{sec-intro}
%%%%%%%%%%%%%%%%%%%%%%%%%%%%%%%%%%%%%%%%%%%%%%%%%%%%%%%%

Relativistic hydrodynamics is successful in describing the evolution of quark-gluon plasma (QGP) produced in ultrarelativistic heavy-ion collisions from the very early times of a fraction of ${\rm fm}/c$ after the collision up to freeze-out at about 10 ${\rm fm}/c$, see the review articles \cite{Heinz:2013th,Gale:2013da,Florkowski:2017olj}. This success is surprising as hydrodynamics is usually applied to systems in local thermodynamic equilibrium while the QGP from the earliest phase of ultrarelativistic heavy-ion collisions is expected to be far from equilibrium and strongly anisotropic. The puzzle of why hydrodynamic-like behaviour occurs in the non-equilibrium system from the early phase of relativistic heavy-ion collisions can be addressed in different ways. 

One possible explanation of the early onset of hydrodynamics is based on the finding that non-equilibrium systems tend to evolve along a hydrodynamic attractor trajectory, see \cite{Heller:2015dha} and the review \cite{Jankowski:2023fdz}. Those which are initially far from the attractor approach it rapidly when the anisotropy is still high, and then evolve along the hydrodynamic attractor.

Another explanation of the success of the hydrodynamic approach uses the observation that the assumption of pressure isotropy can be partially relaxed and the system under consideration still behaves as a fluid. This observation led to the formulation of anisotropic hydrodynamics, see \cite{Florkowski:2010cf,Martinez:2010sc} and the review \cite{Strickland:2014pga}. One takes into account that the pressure in the direction transverse to the collision axis, $p_T$, is usually much larger than the longitudinal pressure, $p_L$, at the early stage of relativistic heavy-ion collisions. The energy-momentum tensor is characterized not by a single pressure $p$, as in case of isotropic systems, but by $p_T$ and $p_L$. The equations of hydrodynamics which are provided by the continuity equation of the energy-momentum tensor need to be supplemented not by one but two extra constraints to close the system of equations. In this way the regime of applicability of a hydrodynamic treatment is extended to non-equilibrium systems with strongly anisotropic momentum distributions.

The authors of the study \cite{Vredevoogd:2008id} found that as long as a system is boost-invariant, and its energy-momentum tensor is traceless and initially diagonal, the system at the early stage of its evolution behaves in some respects as a fluid. A similar finding was presented in \cite{Carrington:2021qvi,Carrington:2023nty,Carrington:2024vpf}, see also Sec.~4.5.2 of Ref.~\cite{Romatschke:2017ejr}, where the glasma from the earliest stage of ultrarelativistic heavy-ion collisions was studied. Although the system is far from equilibrium and its dynamics is governed by the Yang-Mills equations,  glasma temporal evolution resembles that of hydrodynamics. Specifically, an approximate eccentricity scaling of the glasma elliptic flow was found \cite{Carrington:2021qvi,Carrington:2023nty,Carrington:2024vpf}. 

The concept of glasma belongs to the Color Glass Condensate effective theory (see, for example, the review \cite{Gelis:2010nm}) which is based on a separation of scales between hard valence partons and soft gluons. The glasma is the system of coherent chromodynamic fields generated by the valence partons of incoming nuclei. The soft gluons are approximated as classical fields while the valence partons, which carry coloured charges, are classical sources. The dynamics of glasma fields is determined by the classical Yang-Mills equations with sources provided by the valence partons. Observables expressed through chromodynamic fields are obtained by averaging over a Gaussian distribution of colour charges within each incoming nucleus. 

Various characteristics of the glasma have been studied for over two decades using more and more advanced numerical simulations, see Refs.~\cite{Sun:2019fud,Boguslavski:2021buh,Ipp:2021lwz,Avramescu:2023qvv} as examples of recent works in this direction. There are also analytic approaches but usually of limited applicability. In our study we use a method specially designed to describe the earliest phase of relativistic heavy-ion collisions. The method is sometimes called a `near field expansion.' It was proposed in \cite{Fries:2005yc} and further developed in \cite{Fukushima:2007yk,Fujii:2008km,Chen:2015wia,Fries:2017ina,Carrington:2021qvi,Carrington:2023nty,Carrington:2020ssh,Carrington:2020sww,Carrington:2022bnv,Carrington:2021dvw,Carrington:2024vpf}. The method is based on an expansion of chromodynamic potentials in powers of the proper time, $\tau$, which is treated as a small parameter. The expansion allows one to solve the Yang-Mills equations iteratively.  The results provided by the method are limited to small values of $\tau$ but they are analytic and free of  numerical artifacts like those caused by taking a continuous limit in lattice calculations. 

In the works \cite{Carrington:2021qvi,Carrington:2023nty,Carrington:2024vpf} by two of us the energy-momentum tensor of glasma produced in Pb-Pb and Pb-Ca collisions was calculated. The beginning of the process of the system's equilibration is clearly seen by analyzing the temporal evolution of the energy density, transverse and longitudinal pressures. The glasma collective flow characterized by the Fourier coefficient $v_2$ was calculated by including the transverse Poynting vectors in the analysis. It was found that $v_2$ is correlated with the spatial eccentricity of the system similarly as in hydrodynamics. The aim of this paper is to study in more detail the extent to which the glasma behaves as a fluid and to clarify the physical origins of this behaviour.

Throughout the paper we use the natural system of units with $c = \hbar = k_B =1$. Greek letters $\mu, \nu, \rho, \dots$ label components of four-vectors. Vectors transverse to the collision axis $z$ are denoted $\vec x_\perp$ and their components are indexed with Latin letters $i,j,k, \dots$. Latin letters  from the beginning of the alphabet $a,b, c \dots$ label colour components of elements of the SU($N_c$) gauge group in the adjoint representation. We use three systems of coordinates: Minkowski $(t, z, x, y)$, light-cone $(x^+,x^-, x, y)$ and Milne $(\tau, \eta, x, y)$, where $x^\pm = (t\pm z)/\sqrt{2}$, $\tau=\sqrt{t^2-z^2}$ and $\eta=\ln(x^+/x^-)/2$. The indices of vector and tensor components, like $A^t$, $F^{+x}$, $T^{\tau \eta}$, clearly show which coordinates are used. 

%%%%%%%%%%%%%%%%%%%%%%%%%%%%%%%%%%%%%%%%%%%%%%%%%%%%%%%%
\section{Summary of computational method}
\label{sec-method}
%%%%%%%%%%%%%%%%%%%%%%%%%%%%%%%%%%%%%%%%%%%%%%%%%%%%%%%%

We summarize below the method to calculate the energy-momentum tensor of the glasma from early stage of relativistic heavy-ion collisions using a proper time expansion. More details can be found in \cite{Carrington:2021qvi,Carrington:2023nty,Carrington:2020ssh}. 

We consider a collision of two heavy ions moving with the speed of light towards each other along the $z$ axis and colliding at $t=z=0$. The vector potential of the gluon field is described with the ansatz \cite{Kovner:1995ts} 
\ba
\nn
A^+(x) &=& \Theta(x^+)\Theta(x^-) x^+ \alpha(\tau,\vec x_\perp) ,
\\\label{ansatz}
A^-(x) &=& -\Theta(x^+)\Theta(x^-) x^- \alpha(\tau,\vec x_\perp) ,
\\ \nn
A^i(x) &=& \Theta(x^+)\Theta(x^-) \alpha_\perp^i(\tau,\vec x_\perp)
+\Theta(-x^+)\Theta(x^-) \beta_1^i(x^-,\vec x_\perp)
+\Theta(x^+)\Theta(-x^-) \beta_2^i(x^+,\vec x_\perp) ,
\ea
where the functions $\beta_1^i(x^-,\vec x_\perp)$ and $\beta_2^i(x^+,\vec x_\perp)$ represent the pre-collision potentials, and the functions $\alpha(\tau,\vec x_\perp)$ and $\alpha_\perp^i(\tau,\vec x_\perp)$ are the post-collision potentials. 
In the forward light-cone the vector potential satisfies the sourceless Yang-Mills equations but the sources enter through boundary conditions that connect the pre-collision and post-collision potentials. The boundary conditions are
\ba
\label{cond1}
\alpha^{i}_\perp(0,\vec{x}_\perp) &=& \alpha^{i(0)}_\perp(\vec{x}_\perp) 
= \lim_{\text{w}\to 0}\left(\beta^i_1 (x^-,\vec{x}_\perp) + \beta^i_2(x^+,\vec{x}_\perp)\right) ,
\\
\label{cond2}
\alpha(0,\vec{x}_\perp) &=& \alpha^{(0)}(\vec{x}_\perp) 
= -\frac{ig}{2}\lim_{\text{w}\to 0}\;[\beta^i_1 (x^-,\vec{x}_\perp),\beta^i_2 (x^+,\vec{x}_\perp)] ,
\ea
where the notation $\lim_{\text{w}\to 0}$ indicates that the width of the sources across the light-cone is taken to zero, as the colliding nuclei are infinitely contracted. 

We find solutions valid for early post-collision times by expanding the Yang-Mills equations in the proper time $\tau$. Using these solutions we can write the post-collision field-strength tensor, and energy-momentum tensor, in terms of the initial potentials $\alpha(0, \vec x_\perp)$ and $\vec\alpha_\perp(0, \vec x_\perp)$ and their derivatives, which in turn are expressed through the pre-collision potentials $\vec \beta_1(x^-,\vec x_\perp)$ and $\vec \beta_2(x^+,\vec x_\perp)$ and their derivatives. 

The next step is to use the Yang-Mills equations to write the pre-collision potentials in terms of the colour charge distributions of the incoming ions. One then averages over a Gaussian distribution of colour charges within each nucleus. The average of a product of colour charges can be written as a sum of terms that combine the averages of all possible pairs, which is called Wick's theorem. We use the Glasma Graph approximation \cite{Lappi:2017skr} which means that we apply Wick's theorem not to colour charges but to gauge potentials.

The correlator of two pre-collision potentials from different ions is assumed to be zero as the potentials are not correlated to each other. The building blocks of all physical quantities we study are the correlators for two potentials from the same ion
\be
\label{core5-20}
\delta^{ab} B_n^{ij}(\vec{x}_\perp,\vec y_\perp) \equiv 
\lim_{{\rm w} \to 0}  \langle \beta_{n\,a}^i(x^-,\vec x_\perp) \beta_{n\,b}^j(y^-,\vec y_\perp)\rangle  ,
~~~~~~n=1,~\,2
\ee
and their derivatives. The correlators (\ref{core5-20}) are expressed through the colour charge surface density of a given ion $\mu_1(\vec x_\perp)$ or $\mu_2(\vec x_\perp)$, see Sec.~II of Ref.~\cite{Carrington:2021qvi}. These functions are a phenomenological input to our calculations, and we have used a Woods-Saxon distribution projected on the plane transverse to the collision axis. 

Our results are obtained for the SU(3) gauge group, $g=1$, the saturation scale $Q_s = 2$ GeV and infrared cutoff $m=0.2$ GeV. We also use the relation $\bar\mu = Q_s^2/g^4$ where $\bar\mu$ is the value of $\mu_1(\vec x_\perp)$ and $\mu_2(\vec x_\perp)$ at the center of the nucleus. If not stated otherwise the glasma energy-momentum tensor is computed for Pb-Pb collisions at eighth order of the proper time expansion (all contributions up to eighth order are summed). 

%\newpage
%%%%%%%%%%%%%%%%%%%%%%%%%%%%%%%%%%%%%%%%%%%%%%%%%%%%%%%%
\section{Equation of universal flow}
\label{sec-uni-hydro}
%%%%%%%%%%%%%%%%%%%%%%%%%%%%%%%%%%%%%%%%%%%%%%%%%%%%%%%%

In this section we explain why the glasma is expected to behave similarly to a fluid. Our arguments are inspired by the study \cite{Vredevoogd:2008id} where an equation for the flow transverse to the collision axis was derived adopting a few assumptions relevant for the early stage of ultrarelativistic-heavy-ion collisions. We arrive at the same equation, which we call `the equation of universal flow', but our derivation presented in Sec.~\ref{sec-uni-hydro-derivation} is somewhat different. In Sec.~\ref{sec-uni-hydro-order-by-order} we explicitly show that the assumptions used to derive the equation of universal flow are satisfied by the energy-momentum tensor obtained in \cite{Carrington:2021qvi} using an expansion in powers of $\tau$. A similar analysis is given in Sec.~4.5.2 of Ref.~\cite{Romatschke:2017ejr}. We further show that the equation is exactly satisfied order by order by the energy-momentum tensor obtained with a proper time expansion up to seventh order.

%----------------------------------------------------------------------------------------------------------------------
\subsection{Derivation}
\label{sec-uni-hydro-derivation}
%----------------------------------------------------------------------------------------------------------------------

The equations of relativistic hydrodynamics are obtained from the continuity equation of the energy-momentum tensor
\be
\label{cont-EM-tensor}
\nabla_\mu T^{\mu \nu} = \partial_\mu T^{\mu \nu} + \Gamma^\mu_{\mu \rho} T^{\rho \nu} 
+ \Gamma^\nu_{\mu \rho} T^{\mu \rho} = 0,
\ee
where $T^{\mu \nu}$ is the energy-momentum tensor and $\nabla_\mu$ is the covariant derivative which in the case of curvilinear coordinates includes Christoffel symbols, denoted $\Gamma$. One obtains the hydrodynamic equations assuming a specific structure of the tensor $T^{\mu \nu}$, which in case of ideal hydrodynamics is
\be
\label{EM-tensor-ideal-hydro}
T^{\mu \nu}_{\rm ideal}  = ({\cal E} + {\cal P}) u^\mu u^\nu - {\cal P} \, g^{\mu \nu} ,
\ee
where ${\cal E}$ is the energy density, ${\cal P}$ is the pressure, $u^\mu$ is the hydrodynamic four-velocity normalized as $u^\mu u_\mu =1$, and $g^{\mu \nu}$ is the metric tensor. Since the four-velocity is normalized, there are five functions which enter the continuity equation (\ref{cont-EM-tensor}) and one needs an extra equation to close the system. One usually adds an equation of state which relates the energy density to the pressure.  In case of a conformally invariant system, the energy-momentum tensor is traceless and the ideal fluid form (\ref{EM-tensor-ideal-hydro}) provides ${\cal E} = 3{\cal P}$. 

There are three facts that are needed to derive the equation of universal flow.

\begin{itemize}

\item The glasma evolution is studied only in a very short time interval after its formation. 

\item The glasma energy-momentum tensor is initially diagonal.

\item The glasma is invariant under Lorentz boosts along the collision axis.

\end{itemize}
In Minkowski coordinates the initial energy-momentum tensor is diagonal and $T^{tt} = T^{xx} = T^{yy} = - T^{zz} = {\cal E}_0$. One expects that the diagonal components of the energy-momentum tensor remain much bigger than the off-diagonal ones at very early times. Since the glasma is boost invariant, the energy-momentum tensor in Milne coordinates is independent of the space-time rapidity $\eta$. The continuity equation (\ref{cont-EM-tensor}) in Milne coordinates and with $\nu=x$ provides 
\be
\label{cont-eq-x-1}
\Big(\frac{\partial}{\partial \tau} + \frac{1}{\tau}\Big) T^{\tau x} 
+ \frac{\partial T^{xx}}{\partial x} + \frac{\partial T^{yx}}{\partial y}  = 0,
\ee
where the vanishing term $\partial T^{\eta x}/\partial \eta$ is omitted. 

We write the energy-momentum tensor in Milne coordinates in terms of the components of the tensor in Minkowski coordinates. Since we are interested in the glasma in the mid-rapidity region we set $\eta = 0$. One finds
\ba
\label{EM-tensor-Milne-gen}
T_{\rm Milne}^{\mu \nu} = 
\left(\begin{array}{cccc}
T^{tt}  & \frac{1}{\tau}T^{tz}& T^{tx} &  T^{ty} \\
\frac{1}{\tau} T^{zt} & \frac{1}{\tau^2} T^{zz}   &  \frac{1}{\tau} T^{zx} & \frac{1}{\tau} T^{zy} \\
T^{xt} & \frac{1}{\tau}T^{xz} & T^{xx} &   T^{xy} \\
T^{yt} & \frac{1}{\tau} T^{yz} & T^{yx} & T^{yy} 
\end{array}\right) .
\ea
Equation (\ref{EM-tensor-Milne-gen}) shows that at $\eta=0$ the components of the energy-momentum tensor that enter Eq.~(\ref{cont-eq-x-1}) coincide with their counterparts in Minkowski coordinates up to factors that are powers of $\tau$ (the components $T^{xx}, T^{xy}, T^{yx}, T^{yy}$ are the same in both coordinate systems for any $\eta$). We also observe that for $\eta = 0$, we have $z=0$ and $\tau = t$. Replacing the proper time $\tau$ by $t$ and $T^{\tau x}$ by $T^{t x}$, Eq.~(\ref{cont-eq-x-1}) becomes
\be
\label{cont-eq-x-2}
\Big(\frac{\partial}{\partial t} + \frac{1}{t}\Big) T^{tx} + \frac{\partial T^{xx}}{\partial x}  = 0 
\ee
where we have ignored the term that depends on $T^{yx}$ which is much smaller than  $T^{xx}$ at early times (see Eq.~(\ref{Tmilne-orders}) and the explanation below). 
We will assume that $T^{xx}$ weakly depends on time (see again Eq.~(\ref{Tmilne-orders}) and the explanation below), which means that equation (\ref{cont-eq-x-2}) is solved by
\be
T^{tx} = - \frac{1}{2} \, t \frac{\partial T^{xx}}{\partial x} + \frac{C}{t} ,
\ee
where $C$ is an arbitrary constant. Since $T^{tx} = 0$ at $t = 0$ we obtain the equation
\be
\label{uni-hydro-xx}
T^{tx} = - \frac{1}{2} \, t \, \frac{\partial T^{xx}}{\partial x} ,
\ee
which tells us that the flow is generated by the gradient of pressure, similarly to the Euler equation.

\begin{figure}[t]
\begin{center}
\includegraphics[width=12cm]{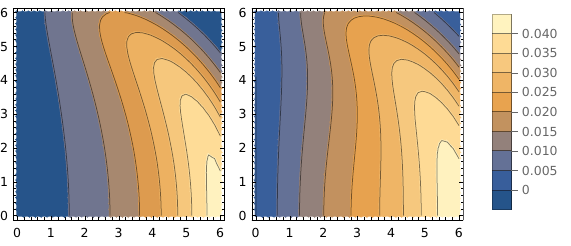}
\end{center}
\vspace{-8mm}
\caption{The left and right sides of Eq.~(\ref{uni-hydro-xx}) for $t=0.06$~fm, $\eta=0$ and $b=2$~fm. Contour plots cover  the upper-right quadrant of the transverse plane and, the axes show the $x$ and $y$ coordinates in fm. Values of components of the energy-momentum tensors are in ${\rm GeV/fm^3}$.}
\label{fig-flow-grad}
\end{figure}

In Refs.~\cite{Carrington:2021qvi,Carrington:2023nty} we obtained the energy-momentum tensor of the glasma from Pb-Pb collisions using a proper time expansion of chromodynamic potentials. In Fig.~\ref{fig-flow-grad} we show the left and right sides of Eq.~(\ref{uni-hydro-xx}) as contour plots in the transverse $x$-$y$ plane, for $t = 0.06$~fm, $\eta = 0$ and impact parameter $b=2$ fm. Fig.~\ref{fig-flow-grad} clearly shows that Eq.~(\ref{uni-hydro-xx}) is satisfied with good accuracy. To obtain a quantitative measure we calculate
\be
\Delta = \frac{|L|-|R|}{|L|+|R|} = 0.184 \nonumber
\ee
where $|L|$ and $|R|$ indicate the average over the transverse plane of the absolute value of the left and right sides of equation (\ref{uni-hydro-xx}).

Initially $T^{tt} = T^{xx}$ and approximately equality is expected to hold at later times that are not too large. Consequently, Eq.~(\ref{uni-hydro-xx}) can be written as 
\be
\label{uni-hydro-tt}
T^{tx} \approx - \frac{1}{2} \, t \, \frac{\partial T^{tt}}{\partial x} 
\ee
which is the equation of universal flow first obtained in \cite{Vredevoogd:2008id}. There is an analogous equation with $x$ replaced by $y$. In the subsequent section we will show that it is solved exactly order by order in the proper time expansion. Although the derivation suggests that Eq.~(\ref{uni-hydro-tt}) is an approximation to Eq.~(\ref{uni-hydro-xx}), in fact (\ref{uni-hydro-tt}) is more precise. 

%----------------------------------------------------------------------------------------------------------------------
\subsection{Order by order analysis}
\label{sec-uni-hydro-order-by-order}
%----------------------------------------------------------------------------------------------------------------------

Keeping terms of order not higher than $\tau^1$, the energy-momentum tensor of the glasma computed in Refs.~\cite{Carrington:2021qvi,Carrington:2023nty} has the form
\ba
T^{\mu\nu}_{\rm Milne} = \left(
\begin{array}{cccc}
T^{\tau \tau}_0 ~~~ & ~~~ \tau T^{\tau\eta}_1 ~~~ & ~~~ \tau T^{\tau x}_1 ~~~ & ~~~ \tau T^{\tau y}_1 \\
\tau T^{\eta \tau}_1 & \frac{1}{\tau^2} T^{\eta\eta}_{-2}+  T^{\eta\eta}_{0} &  T^{\eta x}_0 & T^{\eta y}_0 \\
\tau T^{x \tau}_1 & T^{x\eta}_{0}  &  T^{xx}_0 & 0 \\
\tau T^{y \tau}_1 & T^{y\eta}_{0}  &  0 & T^{yy}_0 \\
\end{array}
\right),
\label{Tmilne-orders}
\ea
where $T^{\mu \nu}_n$ is the coefficient of term in $T^{\mu \nu}$ that is proportional to $\tau^n$. We note that the term proportional to $\tau^{-2}$ in $T^{\eta \eta}$ appears because the tensor is written in Milne coordinates.

Equation (\ref{Tmilne-orders}) shows that the off-diagonal component $T^{yx}$ is higher order in $\tau$ than the diagonal components $T^{xx}$ and $T^{yy}$, which allows us to drop this term when we go from Eq.~(\ref{cont-eq-x-1}) to Eq.~(\ref{cont-eq-x-2}).  One also sees that the leading contribution to $T^{xx}$ is time independent, which justifies the assumption we made to solve Eq.~(\ref{cont-eq-x-2}) and get the solution (\ref{uni-hydro-xx}). Since $T^{\tau \tau}_0 = T^{xx}_0 = T^{yy}_0 =-T^{\eta\eta}_{-2} = {\cal E}_0$ and at mid-rapidity $T^{\tau\tau}=T^{tt}, ~T^{\tau x}=T^{tx}$ and $\tau=t$, Eq.~(\ref{uni-hydro-xx}) gives
\be
\label{uni-hydro-tt-n=0}
T^{tx}_1 = - \frac{1}{2} \frac{\partial T^{tt}_0}{\partial x} .
\ee
Thus we see that the equation of universal  flow (\ref{uni-hydro-tt}), which we argued is approximate at early times, is satisfied exactly at lowest order in the proper time expansion. 

\begin{figure}[t]
\begin{center}
\includegraphics[width=14cm]{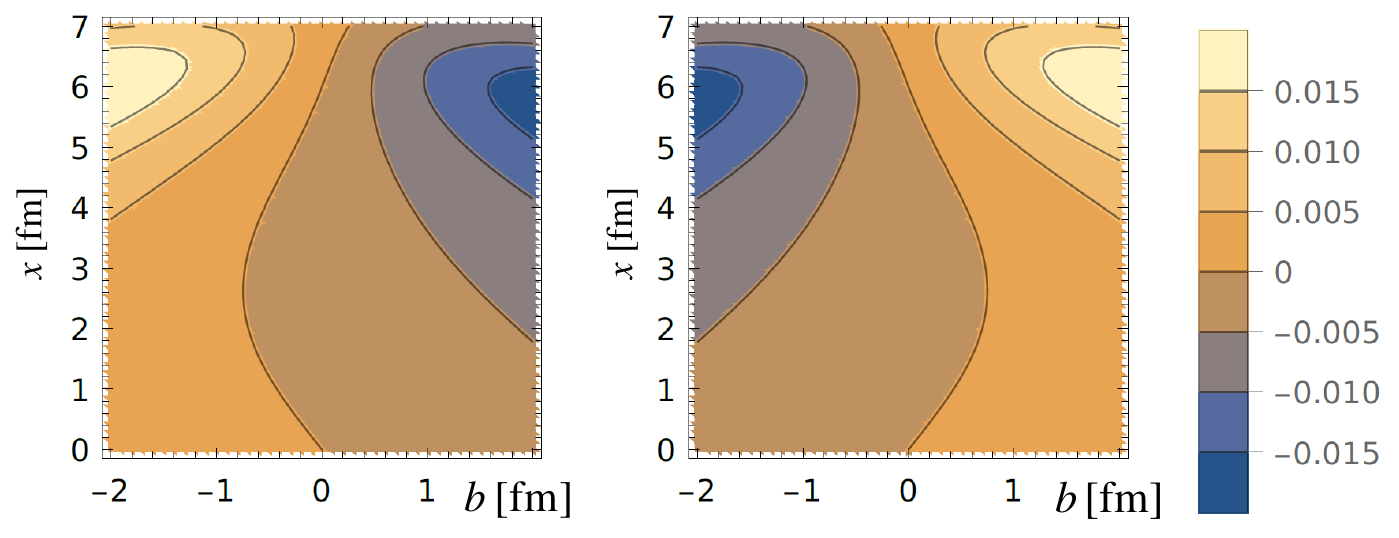}
\end{center}
\vspace{-8mm}
\caption{The differences of the two sides of (\ref{uni-hydro-tt}) as a function of $x$ and $b$ at $y=0$ and $\tau = 0.04$ fm at seventh order in the proper time expansion. The left panel is $\eta=0.5$ and the right is $\eta=-0.5$.}
\label{fig-plots-eta-non-zero} 
\end{figure}

The natural generalization of Eq.~(\ref{uni-hydro-tt-n=0}) to higher orders in the proper time expansion is
\be
\label{uni-hydro-n} 
T^{t x}_{n+1} = - \frac{1}{2}  \frac{\partial T^{tt}_n}{\partial x} .
\ee
Using the explicit expressions of $T^{t x}_{n+1}$ and $T^{tt}_n$ for $n=0,1, \dots 6, 7$ obtained in \cite{Carrington:2021qvi,Carrington:2023nty}, we have checked that Eq.~(\ref{uni-hydro-n}) is exactly satisfied by these coefficients at $\eta=0$. It is reasonable to expect that Eq.~(\ref{uni-hydro-n}) is satisfied at mid-rapidity for any order $n$ in the proper time expansion. In this case the equation of universal flow (\ref{uni-hydro-tt}) for glasma is not approximate but valid not only for short times but for any time at mid-rapidity. 

When values of $\eta$ are small but non-zero the violation of Eq.~(\ref{uni-hydro-n}) is small. For a  symmetric collision of equal size ions with centers displaced symmetrically in the $x$ direction by $\pm b/2$, the violation for $\eta\ne 0$  is odd under interchange of the two ions. It is even under $(\eta,\vec b)\to -(\eta,\vec b)$ and at $\vec R=0$ it is odd under $\vec b \to -\vec b$. Figure ~\ref{fig-plots-eta-non-zero} shows contour plots of the differences of the two sides of Eq.~(\ref{uni-hydro-tt}) as a function of $x$ and $b$ for $y=0$ and $\eta=\pm 0.5$. The energy-momentum tensor is computed at $\tau = 0.04$ fm at seventh order in the proper time expansion. 

%%%%%%%%%%%%%%%%%%%%%%%%%%%%%%%%%%%%%%%%%%%%%%%%%%%%%%%%
\section{Glasma as ideal or anisotropic fluid}
\label{sec-ideal-and-anisotropic-hydro}
%%%%%%%%%%%%%%%%%%%%%%%%%%%%%%%%%%%%%%%%%%%%%%%%%%%%%%%%

In Sec.~\ref{sec-uni-hydro} we showed that the glasma energy-momentum tensor satisfies the equation of universal flow (\ref{uni-hydro-tt}) and we therefore expect that, at least in some ways, it will behave like a fluid. In this section we investigate the extent to which the glasma energy-momentum tensor can be modelled with a hydrodynamic form. 

If the glasma were an ideal fluid its energy-momentum tensor would be given by Eq.~(\ref{EM-tensor-ideal-hydro}). Since the glasma energy-momentum tensor is traceless $T^\mu_{\;\; \mu} = 0$, one gets the equation of state ${\cal E} = 3{\cal P}$ and the energy-momentum tensor (\ref{EM-tensor-ideal-hydro}) becomes
\be
\label{EM-tensor-ideal-hydro-massless}
T^{\mu \nu}_{\rm ideal}  = \frac{4}{3} \,{\cal E} u^\mu u^\nu - \frac{1}{3}\,{\cal E}\, g^{\mu \nu} .
\ee

Since the glasma initial longitudinal pressure is negative, we expect that anisotropic hydrodynamics \cite{Florkowski:2010cf,Martinez:2010sc} will provide a better description. We therefore consider 
\be
\label{EM-tensor-anisotropic-hydro}
T^{\mu \nu}_{\rm aniso}  = ({\cal E} + {\cal P}_T ) u^\mu u^\nu 
- {\cal P}_T g^{\mu \nu} - ({\cal P}_T- {\cal P}_L) z^\mu z^\nu,
\ee
where ${\cal P}_T$ and ${\cal P}_L$ are the transverse and longitudinal pressures and the four-vector $z^\mu$ defines the direction of the longitudinal pressure. It is space-like, orthogonal to the fluid four-velocity $(u^\mu z_\mu = 0)$ and normalized as  $z^\mu z_\mu = -1$. In the local rest frame of the fluid element we have $u^\mu = (1, 0, 0, 0)$ and $z^\mu = (0, 1, 0, 0)$. The energy-momentum tensor is diagonal and of the form $T^{\mu\nu} = {\rm diag}({\cal E}, {\cal P}_L, {\cal P}_T, {\cal P}_T)$. Due to the tracelessness of the glasma energy-momentum tensor, the equation of state is ${\cal E} = 2{\cal P}_T + {\cal P}_L$.  

Although a dissipative contribution is not explicitly included in Eq.~(\ref{EM-tensor-anisotropic-hydro}), the corresponding hydrodynamics \cite{Florkowski:2010cf,Martinez:2010sc} is not ideal. Entropy production proportional to ${\cal P}_T- {\cal P}_L$ is postulated, which provides an extra equation that is used to close the system of hydrodynamic equations. When ${\cal P}_T$ and ${\cal P}_L$ become equal to each other, the entropy production ceases and anisotropic hydrodynamics changes into the ideal version. 

Our aim is to assess how well the glasma energy-momentum tensor $T^{\mu \nu}_{\rm glasma}$ can be represented by the tensor (\ref{EM-tensor-ideal-hydro-massless}) or (\ref{EM-tensor-anisotropic-hydro}). For this purpose we calculate $T^{\mu \nu}_{\rm glasma}$ at a large set of grid points in the transverse $x$-$y$ plane. At each point we find the parameters ${\cal E}$, ${\cal P}_T$, ${\cal P}_L$, $u^\mu$ and $z^\mu$ which enter either the tensors (\ref{EM-tensor-ideal-hydro-massless}) or (\ref{EM-tensor-anisotropic-hydro}) as follows. We find the eigenvalues $\lambda$ and eigenvectors $v^\mu$ of the equation $T_{\rm glasma}^{\mu\nu} v_\nu = \lambda v^\mu$. Since the energy-momentum tensor is a $4 \times 4$ matrix that is symmetric and real it has four eigenvectors and four real eigenvalues. One eigenvector is time-like and the remaining three are space-like. The eigenvalue corresponding to the time-like eigenvector is identified as the energy density and the eigenvector is the fluid four-velocity. We note that $T_{\rm ideal}^{\mu\nu}  u_\nu =T_{\rm aniso}^{\mu\nu} u_\nu = {\cal E} u^\mu$. The negative eigenvalue is identified as ${\cal P}_L$. The transverse pressure is then ${\cal P}_T= ({\cal E}-{\cal P}_L)/2$. The space-like eigenvector $z^\mu$ could be obtained in two different ways. We could define it to be the space-like eigenvector that corresponds to the negative eigenvalue. 
Alternatively we could find $z^\mu$ by Lorentz transforming the corresponding vector in the local rest frame $z^\mu_{\rm lrf}=(0,1,0,0)$ to the reference frame moving with the fluid four-velocity. Both of these methods give similar results.

We have applied the procedure described above to construct $T^{\mu\nu}_{\rm ideal}$ and $T^{\mu\nu}_{\rm aniso}$ from the glasma energy-momentum tensor  $T^{\mu\nu}_{\rm glasma}$  computed  using the methods developed in \cite{Carrington:2021qvi,Carrington:2023nty,Carrington:2020ssh}. We displace the centers of two equal size colliding ions symmetrically about the $x$-axis. The energy-momentum tensor is symmetric which means it is fully determined by the ten components: $T^{tt}, T^{tz}, T^{tx}, T^{ty}, T^{zz}, T^{zx},  T^{zy}, T^{xx}, T^{xy}, T^{yy}$. At $z=0$, the components $T^{tz}$, $T^{zx}$ and $T^{zy}$ are identically zero. For zero impact parameter there is cylindrical symmetry and thus $T^{tx} = T^{ty}$, $T^{zx} = T^{zy}$ and $T^{xx} = T^{yy}$. Consequently, when $b=0$ and $z=0$ the energy-momentum tensor is determined by the 5 non-zero components.

\begin{figure}
\begin{center}
\vspace{-8mm}
\includegraphics[width=14cm]{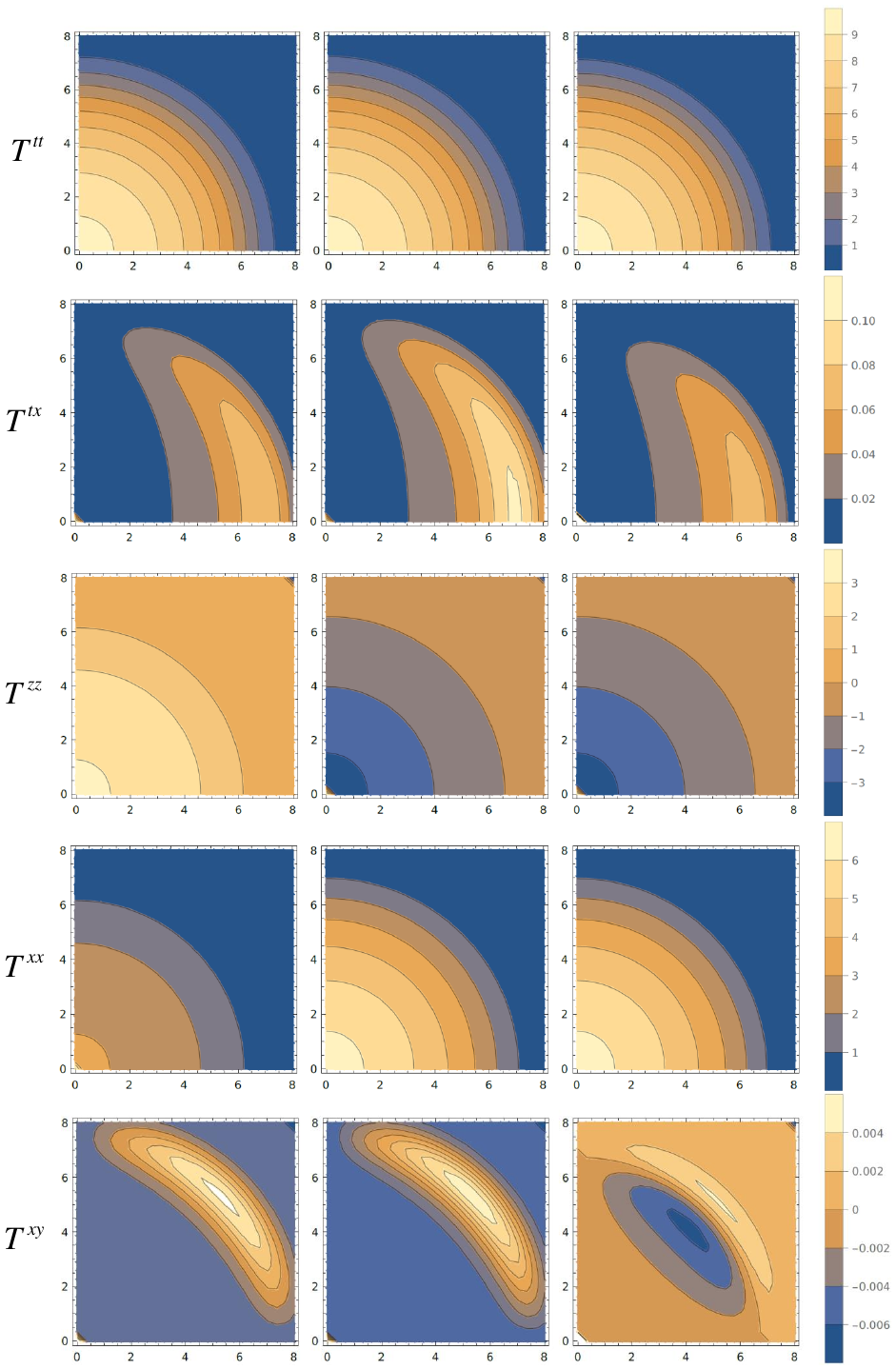}
\end{center}
\vspace{-8mm}
\caption{Components of  $T^{\mu \nu}_{\rm ideal}$ (left column), $T^{\mu \nu}_{\rm aniso}$ (middle column) and $T^{\mu \nu}_{\rm glasma}$ (right column) at $b=0$, $\eta = 0$ and $t=0.06~{\rm fm}$. Contour plots cover the upper-right quadrant of the transverse plane, the axes show $x$ and $y$ coordinates in fm. Values of components of the energy-momentum tensors are in ${\rm GeV/fm^3}$.}
\label{fig-EMT-plots}
\end{figure}

In Fig.~\ref{fig-EMT-plots} we show  contour plots of the non-zero components of the three energy-momentum tensors we have calculated for  impact parameter $b=0$, space-time rapidity $\eta = 0$ and at time $t=0.06~{\rm fm}$. The left, middle and right columns show  $T^{\mu\nu}_{\rm ideal}$, $T^{\mu\nu}_{\rm aniso}$  and $T^{\mu\nu}_{\rm glasma}$, respectively. The contour plots cover one quadrant of the transverse plane and the axes show $x$ and $y$ coordinates in~fm. Values of components of the energy-momentum tensors are given in ${\rm GeV/fm^3}$. The energy-momentum tensor of ideal hydrodynamics represents the $tt$, $tx$ and $xx$ components of the glasma tensor very well, but it badly fails for $zz$ and $xy$. In case of the $zz$ component this is of no surprise as the negative value $T^{zz}_{\rm glasma}$ is in evident conflict with the assumption of isotropic pressure adopted in ideal hydrodynamics. Anisotropic hydrodynamics, which is designed to deal with anisotropic systems, works better. The contour plots of all components of $T^{\mu\nu}_{\rm aniso}$ are very similar to those of $T^{\mu\nu}_{\rm glasma}$, except for the $xy$ component. It should be noted however that the value of this component is much smaller than the other non-zero components.

%\newpage

To quantify the difference between the glasma energy-momentum tensor $T^{\mu\nu}_{\rm glasma}$ and different hydrodynamic formulations we look at the measure
\be
\label{sigma-def}
\sigma^{\mu\nu}_{\rm hydro} = \sqrt{\frac{\sum_j \big(T^{\mu\nu}_{\rm glasma}(\vec{x}_\perp^{\,j}) 
- T^{\mu\nu}_{\rm hydro}(\vec{x}_\perp^{\,j}) \big)^2}
{\sum_j \big(T^{\mu\nu}_{\rm glasma}(\vec{x}_\perp^{\,j}) 
+ T^{\mu\nu}_{\rm hydro}(\vec{x}_\perp^{\,j})\big)^2}} ~ ,
\ee
where the sums are over all points in the transverse plane at which the tensors are computed. The measure $\sigma^{\mu\nu}_{\rm hydro}$ multiplied by a factor of 2 gives the relative difference between $T^{\mu\nu}_{\rm glasma}$ and $T^{\mu\nu}_{\rm hydro}$. If a component vanishes for both $T^{\mu\nu}_{\rm glasma}$ and $T^{\mu\nu}_{\rm hydro}$ the measure (\ref{sigma-def}) is ill defined. In such a case the components of $T^{\mu\nu}_{\rm glasma}$ and $T^{\mu\nu}_{\rm hydro}$ exactly agree with each other and we just set $\sigma^{\mu\nu}$ to zero. 

When $b=0$, $\eta = 0$ and $t=0.06~{\rm fm}$ the measures (\ref{sigma-def}) are 
\ba
\label{sigma-b}
\sigma^{\mu\nu}_{\rm ideal} = \left(
\begin{array}{cccc}
 0.003 & 0 & 0.658 & 0.658 \\
 0 & 18.9 & 0 & 0 \\
 0.658 & 0 & 0.328 & 0.968 \\
 0.658 & 0 & 0.968 & 0.328 \\
\end{array}
\right),
~~~
\sigma^{\mu\nu}_{\rm aniso} = \left(
\begin{array}{cccc}
 0.004 & 0 & 0.727 & 0.727 \\
 0 & 0 & 0 & 0 \\
 0.727 & 0 & 0.003 & 0.973 \\
 0.727 & 0 & 0.973 & 0.003 \\
\end{array}
\right). 
\ea
These results show clearly that ideal hydrodynamics cannot describe the negative value of $T^{zz}_{\rm glasma}$ while anisotropic hydrodynamics represents fairly well all components of the glasma energy-momentum tensor. The values of $\sigma^{\mu\nu}_{\rm hydro}$ averaged over all components are $\langle\sigma^{\mu\nu}_{\rm ideal}\rangle = 1.51$ and $\langle\sigma^{\mu\nu}_{\rm aniso}\rangle = 0.304$.

We can also consider collisions with non-zero impact parameter and non-zero values of spatial rapidity. 
When $b=2$ fm, $\eta = 0$ and $t=0.06~{\rm fm}$ the measures (\ref{sigma-def}) are 
\ba
\label{sigma-e}
\sigma^{\mu\nu}_{\rm ideal} = \left(
\begin{array}{cccc}
 0.003 & 0.922 & 0.648 & 0.647 \\
 0.922 & 19.1 & 0.979 & 0.959 \\
 0.648 & 0.979 & 0.328 & 0.809 \\
 0.647 & 0.959 & 0.809 & 0.328 \\
\end{array}
\right),
~~~
\sigma^{\mu\nu}_{\rm aniso} = \left(
\begin{array}{cccc}
 0.004 & 0.912 & 0.716 & 0.716 \\
 0.912 & 0.000 & 0.0122 & 0.022 \\
 0.716 & 0.012 & 0.003 & 0.849 \\
 0.716 & 0.022 & 0.849 & 0.003 \\
\end{array}
\right). 
\ea
The average values of $\sigma^{\mu\nu}_{\rm hydro}$ are $\langle\sigma^{\mu\nu}_{\rm ideal}\rangle = 1.86$ and $\langle\sigma^{\mu\nu}_{\rm aniso}\rangle = 0.404$. When $b=2$ fm, $\eta = 0.5$ and $t=0.06~{\rm fm}$ the measures (\ref{sigma-def}) are 
\ba
\label{sigma-c}
\sigma^{\mu\nu}_{\rm ideal} = \left(
\begin{array}{cccc}
 0.199 & 0.138 & 0.544 & 0.560 \\
 0.138 & 1.58  & 0.752 & 0.442 \\
 0.544 & 0.752 & 0.208 & 0.666 \\
 0.560 & 0.442 & 0.666 & 0.208 \\
\end{array}
\right),
~~~
\sigma^{\mu\nu}_{\rm aniso} = \left(
\begin{array}{cccc}
 0.080 & 0.402 & 0.616 & 0.625 \\
 0.402 & 0.074 & 0.743 & 0.543 \\
 0.616 & 0.743 & 0.079 & 0.701 \\
 0.625 & 0.543 & 0.701 & 0.079 \\
\end{array}
\right). 
\ea
The average values of $\sigma^{\mu\nu}_{\rm hydro}$ are $\langle\sigma^{\mu\nu}_{\rm ideal}\rangle = 0.519$ and $\langle\sigma^{\mu\nu}_{\rm aniso}\rangle = 0.473$. 

In all cases ideal hydrodynamics fails to accurately reproduce the $zz$ component of the energy-momentum tensor and consequently gives a larger relative error when compared with the CGC result. 
\begin{table}[t]
\centering 
\begin{tabular}{| c | c | c | c |}  
\hline                     
~~ $b$~[fm] ~~ & ~~ $\eta$ ~~ & 
~~ $\langle \tilde{\sigma}^{\mu\nu}_{\rm ideal}\rangle$ 
~~ &~~ $\langle \tilde{\sigma}^{\mu\nu}_{\rm aniso}\rangle$ ~~ \\
\hline
 0   & 0  & 0.0436 & 0.0042 \\
 2  & 0   & 0.0438 & 0.0042 \\
 2 & 0.1 & 0.0433 & 0.0085 \\
 2 & 0.5 & 0.0417 & 0.0395 \\
\hline
\end{tabular}
\caption{The differences of  $T^{\mu\nu}_{\rm glasma}$ and $T^{\mu\nu}_{\rm hydro}$ quantified by the measure (\ref{sigma-tilde-def}). The energy-momentum tensors are computed at $t =0.06$~fm.}
\label{sigma-res}
\end{table}

The measure (\ref{sigma-def}) provides the relative error of the two tensors $T^{\mu\nu}_{\rm glasma}$ and $T^{\mu\nu}_{\rm hydro}$ taking into account all tensor components. The components that are numerically very small like $T^{xy}$ can contribute to the sum with equal importance as the diagonal elements that are large. An alternative definition of the measure that weights the different components of the energy-momentum tensor according to their numerical value can be constructed by using the $tt$ component of the energy-momentum tensor in the denominator
\be
\label{sigma-tilde-def}
\tilde{\sigma}^{\mu\nu}_{\rm hydro} = \sqrt{\frac{\sum_j \big(T^{\mu\nu}_{\rm glasma}(\vec{x}_\perp^{\,j}) 
- T^{\mu\nu}_{\rm hydro}(\vec{x}_\perp^{\,j}) \big)^2}
{\sum_j \big(T^{tt}_{\rm glasma}(\vec{x}_\perp^{\,j}) 
+ T^{tt}_{\rm hydro}(\vec{x}_\perp^{\,j})\big)^2}} ~.
\ee
Using this definition we obtain the results shown in Table~\ref{sigma-res}. The energy-momentum tensors are computed at $t =0.06$~fm. Anisotropic hydrodynamics consistently does better than ideal hydrodynamics for all values of $b$ and $\eta$. The improvement is about a factor of 10 when $\eta=0$ but at $\eta=0.5$ both versions of hydrodynamics model the glasma almost equally well. 

\begin{figure}[t]
\begin{center}
\includegraphics[width=16.3cm]{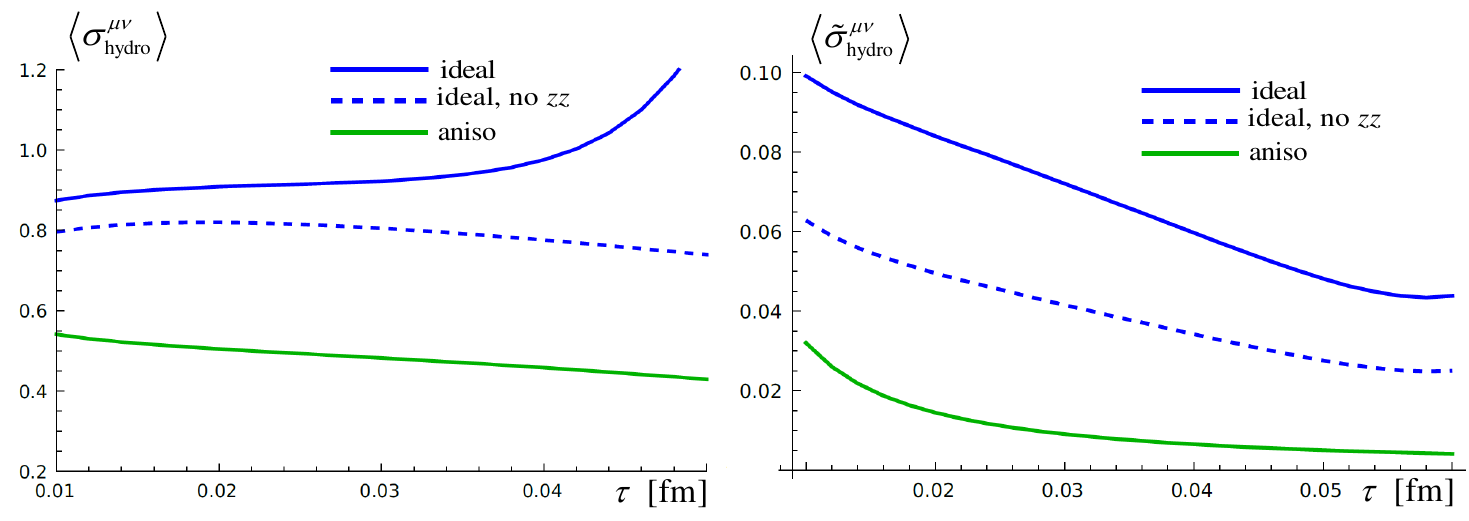}
\end{center}
\vspace{-8mm}
\caption{The average measure (\ref{sigma-def}) (left panel) and (\ref{sigma-tilde-def}) (right panel) for ideal and anisotropic hydrodynamics as a function time for the energy-momentum tensors computed at  $b=1$ fm and $\eta = 0$. The solid blue line represents ideal hydrodynamics and the dotted blue line is without the $zz$ component. The green line is the result for  anisotropic hydrodynamics.}
\label{fig-sigm-vs-time}
\end{figure}

Since the glasma has been found to evolve towards thermodynamic equilibrium \cite{Carrington:2021qvi,Carrington:2023nty}, see also the review \cite{Carrington:2024vpf}, one expects that the hydrodynamic mapping should be closer to the glasma energy-momentum tensor at larger times. To verify this expectation we show in Fig.~\ref{fig-sigm-vs-time} the average measures (\ref{sigma-def}) (left panel) and (\ref{sigma-tilde-def}) (right panel) as a function of time for the energy-momentum tensor at $b=1$ fm and $\eta = 0$. The solid blue line represents ideal hydrodynamics and the dotted blue line is without the $zz$ component which is the most difficult for ideal hydrodynamics to capture. The green line is the result for  anisotropic hydrodynamics. The glasma becomes more and more similar to a fluid as the time advances,  and anisotropic hydrodynamics is a better model than ideal hydrodynamics, or ideal hydrodynamics without the $zz$ component. 

%%%%%%%%%%%%%%%%%%%%%%%%%%%%%%%%%%%%%%%%%%%%%%%%%%%%%%%%
\section{Elliptic flow vs. initial eccentricity}
\label{sec-v2-ecce}
%%%%%%%%%%%%%%%%%%%%%%%%%%%%%%%%%%%%%%%%%%%%%%%%%%%%%%%%

The glasma elliptic flow can be quantified by the Fourier coefficient, $v_2$, and  the spatial eccentricity of the system, $\varepsilon$, defined as \cite{Carrington:2021qvi,Carrington:2023nty,Carrington:2024vpf} 
\be
\label{v2-eccentricity-def}
v_2 = \frac{\int d^2 R \, \frac{(T^{tx})^2 - (T^{ty})^2}{\sqrt{(T^{tx})^2+(T^{ty})^2}}}
{\int d^2 R \, \sqrt{(T^{tx})^2+(T^{ty})^2}}
\text{~~~~~and~~~~~} 
\varepsilon = - \frac{\int d^2 R\,  \frac{R_x^2-R_y^2}{\sqrt{R_x^2+R_y^2}} \,T^{tt}}
{\int d^2 R\, \sqrt{R_x^2+R_y^2} \, T^{tt}} .
\ee
We emphasize that this definition of $v_2$ is not directly related to the elliptic flow studied experimentally in relativistic heavy-ion collisions, see the review \cite{Voloshin:2008dg}. The measured quantity characterizes the asymmetry of the azimuthal distribution of final state particles produced in collisions. Its description in hydrodynamic models requires a proper treatment of the conversion of fluid into particles and the disintegration of the system at a freeze-out hypersurface \cite{Heinz:2013th,Gale:2013da}. Consequently the azimuthal asymmetry of the fluid flow, which is quantified by the coefficient $v_2$ defined by Eq.~(\ref{v2-eccentricity-def}), is significantly washed out by the thermal motion of final state particles. 

The elliptic flow observed in relativistic heavy-ion collisions is zero or almost zero for $b=0$ and grows as $b$ increases \cite{Voloshin:2008dg}. The system's initial eccentricity, $\epsilon(0)$, behaves similarly and consequently the ratio $v_2/\epsilon(0)$ is independent or weakly dependent on impact parameter \cite{Teaney:2010vd}. This is called eccentricity scaling and is generally taken as an indicator of fluid behaviour \cite{Heinz:2013th,Gale:2013da}. The idea is that in hydrodynamics the difference of pressure gradients in and out the reaction plane, which is encoded in the system's spatial eccentricity, is responsible for the azimuthal asymmetry of the transverse collective flow quantified by $v_2$. 

\begin{figure}[t]
\begin{center}
\includegraphics[width=16.3cm]{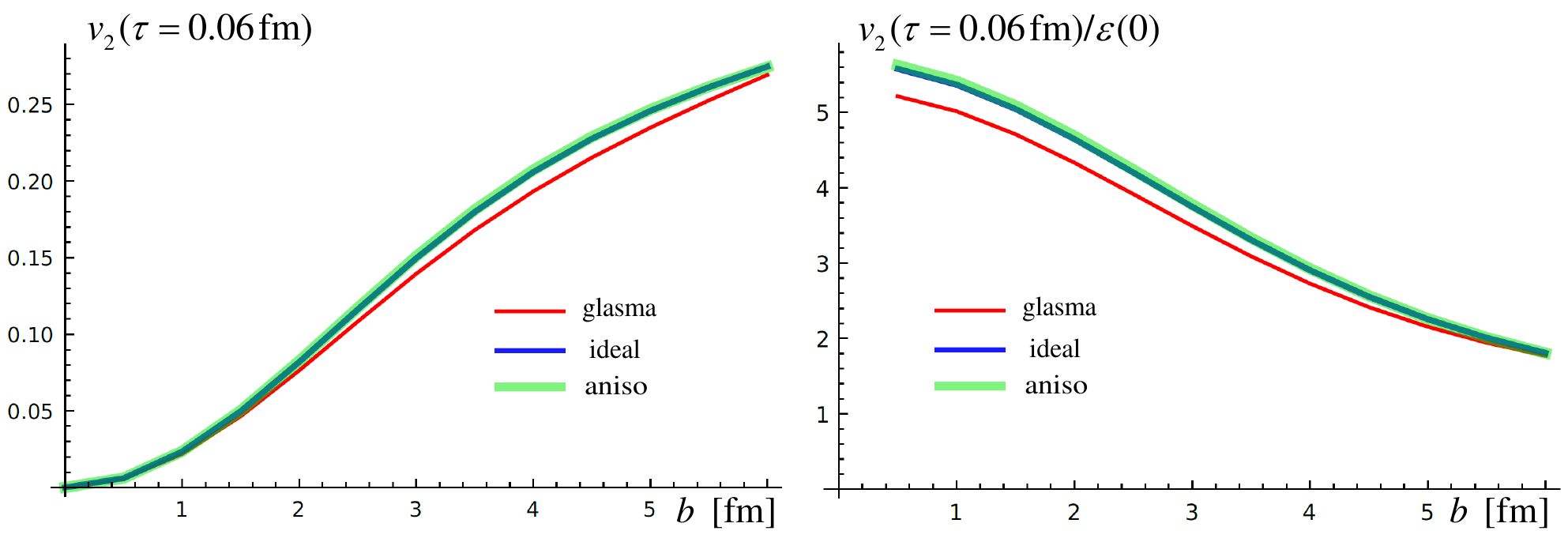}
\end{center}
\vspace{-10mm}
\caption{The elliptic flow coefficient $v_2(t)$ at $t=0.06$~fm and its ratio with the initial eccentricity $\epsilon(0)$. Both quantities are computed directly from the glasma energy-momentum tensor and from the energy-momentum tensor of ideal and anisotropic hydrodynamics. }
\label{plot-dif}
\end{figure}

It was demonstrated in \cite{Carrington:2021qvi,Carrington:2023nty,Carrington:2024vpf} that the glasma elliptic flow $v_2(t)$ coefficient at small but finite time $t$ and the initial eccentricity $\epsilon(0)$ behave in a similar way as functions of impact parameter. The ratio $v_2(t)/\epsilon(0)$ does not show eccentricity scaling but it has a dependence on $b$ which is noticeably weaker than that of $v_2(t)$ and $\epsilon(0)$. Specifically, when the impact parameter grows from 1 to 6 fm, the elliptic coefficients $v_2(t)$ at $t=0.06$~fm increases by the factor of 10 but the ratio $v_2(t)/\epsilon(0)$ decreases only by a factor of 3. This is shown in Fig.~\ref{plot-dif}. The quantities $v_2(t)$ and $\epsilon(0)$ are calculated in three different ways: directly from the glasma energy-momentum tensor and from the energy-momentum tensors of ideal and anisotropic hydrodynamics. Both ideal and anisotropic hydrodynamics give results that are very similar to what is obtained from the glasma. This shows that glasma transverse dynamics can be approximated by hydrodynamics. 

%%%%%%%%%%%%%%%%%%%%%%%%%%%%%%%%%%%%%%%%%%%%%%%%%%%%%%%%
\section{Summary and conclusions}
\label{sec-conclusions}
%%%%%%%%%%%%%%%%%%%%%%%%%%%%%%%%%%%%%%%%%%%%%%%%%%%%%%%%

The evolution of quark-gluon plasma (QGP) produced in ultrarelativistic heavy-ion collisions is well described by relativistic hydrodynamics starting from very early times after the collision when the system is far from equilibrium and highly anisotropic. In refs. \cite{Carrington:2021qvi,Carrington:2023nty} it was argued that this happens because the glasma behaves as fluid even though it is out of equilibrium and its dynamics is governed by the Yang-Mills equation. In this paper we have further explored this problem. 

In the first part of this paper we have discussed the equation of universal flow first obtained in \cite{Vredevoogd:2008id}, which shows that the  transverse collective flow in a given direction is due to the gradient of the energy density of the system in that direction. The equation comes from the continuity of energy-momentum tensor under the assumptions that the system is boost invariant and that the tensor is traceless and initially diagonal. We have rederived the equation using Milne coordinates which greatly simplifies the form of the energy-momentum tensor because it is boost invariant. Using the glasma energy-momentum tensor obtained in a proper time expansion we have also verified the assumptions under which the universal flow equation holds. 

The equation of universal flow has only been derived using a set of assumptions that seem to restrict its validity to early times and systems with weak transverse anisotropy. The most interesting result of the first part of our paper is that the universal flow equation is exactly solved, order by order, by the glasma energy-momentum tensor to seventh order in a proper time expansion. This suggests that the equation is more general than the derivation suggests and that it holds not only for very short times but longer ones as well. 

In the second part of this paper we have analyzed how well the glasma energy-momentum tensor can be mapped onto an energy-momentum tensor of ideal or anisotropic hydrodynamics. We have shown that  ideal hydrodynamics represents the glasma reasonably well with the exception of the $zz$ component. Anisotropic hydrodynamics works very well. The similarity of the results obtained from $T^{\mu \nu}_{\rm glasma}$ and $T^{\mu \nu}_{\rm hydro}$  shows that the hydrodynamic like behaviour of glasma is related to bulk properties of the glasma energy-momentum tensor which survive when the tensor is represented by a hydrodynamic form. Since the glasma evolves towards thermodynamic equilibrium, one expects that the hydrodynamic mapping becomes more accurate as time grows. Our analysis fully confirms this expectation. 

In the last part of this paper we have returned to the primary observation made in \cite{Carrington:2021qvi,Carrington:2023nty,Carrington:2024vpf} that  glasma elliptic flow is correlated with the spatial eccentricity of the system. Both quantities increase with impact parameter, and the ratio of the elliptic flow coefficient at finite proper time divided by the initial eccentricity depends more weakly on impact parameter than $v_2$ itself. We have shown that these features survive when the glasma energy-momentum tensor is represented by either ideal or anisotropic hydrodynamics. 

Our analysis provides a simple and natural explanation of the surprising success of hydrodynamic models of matter produced in ultrarelativistic heavy-ion collisions. These models properly describe a system which can be far from equilibrium because glasma evolution driven by  non-Abelian dynamics strongly resembles hydrodynamic behaviour. The energy-momentum conservation encoded in the continuity equation of energy-momentum tensor, combined with the initial diagonal character of the tensor, its tracelessness, and the boost invariance of the system, play a crucial role. 

Finally, we mention some possible improvements and future directions for this work. One important simplification in our approach is the assumption of boost invariance. Clearly it would be desirable to go beyond this approximation and include some of the effects of the longitudinal dynamics. Unfortunately this would be very difficult because it would require a modification of the ansatz that determines the structure of the gauge potential.

It would be also desirable to check whether the fluid-like behaviour of glasma produced in heavy-ion collisions also occurs in smaller systems obtained from proton-nucleus or proton-proton collisions. Our approach could be easily extrapolated to small mass numbers of colliding nuclei but the extent to which these results would be reliable is unclear. Since our method is fully classical, we assume that the system's dynamics is dominated by long-wavelength modes. This assumption becomes less reliable for smaller systems because the role of short-wavelength quantum modes increases. 

%\newpage
%-----------------------------------------------------------------------
\section*{Acknowledgments}
%-----------------------------------------------------------------------

This work was partially supported by the Natural Sciences and Engineering Research Council of Canada under grant SAPIN-2023-00023. MEC thanks IPhT Saclay for hospitality.

%\newpage

\end{document}